\documentstyle[eqsecnum,aps]{revtex}

\begin{document}
\draft
\title{Universal Formula for the Expectation Value of the Radial Operator under the Aharonov-Bohm
Flux and the Coulomb Field}
\author{W.F. Kao \thanks{%
e-mail: wfgore@cc.nctu.edu.tw}, Y.M. Kao, and D.H. Lin \thanks{%
e-mail: n00ldh00@nchc.gov.tw}}
\address{Institute of Physics, National Chiao Tung University \\
Hsin Chu, Taiwan}
\date{\today}
\maketitle

\begin{abstract}
A useful and universal formula for the expectation value of the
radial operator in the presence of the Aharonov-Bohm flux and the
Coulomb Field is established. We find that the expectation value
$\left\langle r^{\lambda }\right\rangle $ $(-\infty \leq \lambda
\leq \infty )$ is greatly affected due to the non-local effect of
the magnetic flux although the Aharonov-Bohm flux does not have
any dynamical significance in classical mechanics. In particular,
the quantum fluctuation increases in the presence of the magnetic
flux due to the Aharonov-Bohm effect. In addition, the Virial
theory in quantum mechanics is also constructed for the
spherically symmetric system under the Aharonov-Bohm effect.
\end{abstract}



\section{Introduction}

The calculation of the expectation values $\left\langle r^{\lambda
}\right\rangle $ with different power $\lambda $ of the radial
operator $\hat{r}$ is important in the Hellmann-Feynman theory
\cite{1,2}, the interactions of the molecular theory \cite{3}, and
the quantum chemistry \cite{4,4a,4b,4c}. On the other hand, the
Aharonov-Bohm (AB) effect, a topological non-local physical
significance at the quantum level, has shed light on the
understanding of the phenomenon of the fractional quantum Hall
effect \cite{5,6,7,8}, the superconductivity \cite{8,9}, and the
repulsive Bose gases \cite{10} in the last 20 years. It is known
that the global influence of the AB effect affects all charged
particle's systems \cite {11}. In this
paper, we will derive a recurrence formula for the expectation value $%
\left\langle r^{\lambda }\right\rangle $ ($-\infty \leq \lambda \leq \infty $%
) of the radial operator $\hat{r}$ in the presence of the Coulomb
field under the Aharonov-Bohm effect (ABC). Moreover, the Virial
theory in quantum mechanics will also be generalized to the system
with spherical symmetry under the AB flux. The general effect of
the AB flux to the expectation value will also be discussed in
this paper.

The fixed-energy bare Green's function $G^{0}({\bf r,r}^{\prime
};E)$ for a charged particle with mass $m$ propagating from ${\bf r}^{\prime } $
to ${\bf r}$ satisfies the Schr\"{o}dinger equation
\begin{equation}
\left[ E-\hat{H}_{0}({\bf r},\frac{\hbar }{i}{\bf \nabla )}\right] G^{0}(%
{\bf r,r}^{\prime };E)=\delta ^{3}({\bf r-r}^{\prime }).
\label{a1}
\end{equation}
Here the Hamiltonian of the system is given by $\hat{H}_{0}=-$ $\hbar ^{2}{\bf %
\nabla }^{2}/2m+V({\bf r})$. The angular decomposition of the
Green's function can be written as
\begin{equation}
G^{0}({\bf r,r}^{\prime };E)=\sum_{l=0}^{\infty }\sum_{k=-l}^{l}G_{l}^{0}(r%
{\bf ,}r^{\prime };E)Y_{lk}(\theta ,\varphi )Y_{lk}^{\ast }(\theta ^{\prime
},\varphi ^{\prime }),  \label{a2}
\end{equation}
in the spherically symmetric system with $Y_{lk}$ the well-known
spherical harmonics. Hence the left hand side of the Eq.
(\ref{a1}) can be reduced to the following form:
\[
\left\{ E-\sum_{l=0}^{\infty }\sum_{k=-l}^{l}\left[ -\frac{\hbar ^{2}}{2m}%
\left( \frac{d^{2}}{dr^{2}}+\frac{2}{r}\frac{d}{dr}\right) +\frac{%
l(l+1)\hbar ^{2}}{2mr^{2}}\right] -V(r)\right\}
\]
\begin{equation}
\times G_{l}^{0}(r{\bf ,}r^{\prime };E)Y_{lk}(\theta ,\varphi )Y_{lk}^{\ast
}(\theta ^{\prime },\varphi ^{\prime }).  \label{a3}
\end{equation}

\section{Radial wave equation for a charged particle in a magnetic field}

For a charged particle in a magnetic field, the charged Green's
function $G$ is related to the bare Green's function $G^{0}$ by
the following equation:
\begin{equation}
G({\bf r,r}^{\prime };E)=G^{0}({\bf r,r}^{\prime };E)e^{\frac{ie}{\hbar c}%
\int_{{\bf r}^{\prime }}^{{\bf r}}{\bf A}({\bf \tilde{r})\cdot }d{\bf \tilde{%
r}}},  \label{a4}
\end{equation}
with a globally path-dependent nonintegrable phase factor
\cite{14,15} given above. Here the vector potential ${\bf A}({\bf
\tilde{r})}$ is used to represent the contribution from the
magnetic field. For the Aharonov-Bohm magnetic flux under
consideration, the vector potential can be written as
\begin{equation}
{\bf A(x)=}\left\{
\begin{array}{l}
\frac{1}{2}B\rho \hat{e}_{\varphi }\qquad \qquad (\rho <\epsilon ) \\
\frac{1}{2}B\frac{\epsilon ^{2}}{\rho }\hat{e}_{\varphi }=\frac{\Phi }{2\pi
\rho }\hat{e}_{\varphi }\quad (\rho >\epsilon )
\end{array}
\right. .  \label{a5}
\end{equation}
Here the two-dimensional radial length square is defined as $\rho
^{2}=x^{2}+y^{2}$.
Moreover, $\hat{e}_{\varphi }$ is the unit vector of the coordinate $%
\varphi $ and $\epsilon $ is the radius of region where the
magnetic field
exists. Hence the total magnetic flux is given by $\Phi =\pi \epsilon ^{2}B$%
. Note that the associated magnetic field lines are confined
inside a tube, with radius $\epsilon $, along the $z$-axis. Along
the region free of the magnetic field, the path-dependent
nonintegrable phase factor is given by exp\{$-i\mu
_{0}\int_{P}^{\tau }d\tau ^{\prime }\dot{\varphi}(\tau
^{\prime })\}$. Here we have used the subscript $P$ to represent
the
path-dependent nature of the phase factor. In addition, we have denoted $\dot{\varphi}%
(\tau ^{\prime })=d\varphi /d\tau ^{\prime }$. Moreover, $\mu
_{0}=-2eg/\hbar c$ is a dimensionless numerical factor defined by
$\Phi =4\pi g$. The minus sign we adopted is a matter of
convention. According to the discussion in Ref. \cite{15}, only
phase factors with closed-loop contour are considered where the
description of electromagnetic phenomenon are complete. Hence, we
have
\begin{equation}
n=\frac{1}{2\pi }\int_{P}^{\tau }d\tau ^{\prime }\dot{\varphi}(\tau
^{\prime }),  \label{a8}
\end{equation}
with integer values $n$ corresponding to the winding number. The
magnetic interaction is therefore a purely topological phenomenon.
Therefore the nonintegrable phase factor becomes $\exp \{-i\mu
_{0}\left( 2n\pi \right) \}$. With the help of equality between
the associated Legendre polynomial $P_{\nu }^{\mu }(z)$ and the
Jacobi function $P_{n}^{\left( \alpha ,\beta \right) }(z)$
\cite{16,17}, we find that
\begin{equation}
P_{l}^{k}(\cos \theta )=(-1)^{k}\frac{\Gamma (l+k+1)}{\Gamma (l+1)}\left(
\cos \frac{\theta }{2}\sin \frac{\theta }{2}\right) ^{k}P_{l-k}^{\left(
k,k\right) }(\cos \theta ).  \label{a10}
\end{equation}
Therefore the angular part of the Green's function in the
expression (\ref{a3}) can be shown to be
\[
\sum_{k{\bf =-}l}^{l}Y_{lk}(\theta ,\varphi )Y_{lk}^{\ast }(\theta ^{\prime
},\varphi ^{\prime })=\sum_{k{\bf =-}l}^{l}\frac{2l+1}{4\pi }\frac{\Gamma
\left( l-k+1\right) }{\Gamma \left( l+k+1\right) }P_{l}^{k}(\cos \theta
)P_{l}^{k}(\cos \theta ^{\prime })e^{ik(\varphi -\varphi ^{\prime })}
\]
\[
=\sum_{k{\bf =-}l}^{l}\left[ \frac{2l+1}{4\pi }\frac{\Gamma \left(
l-k+1\right) \Gamma \left( l+k+1\right) }{\Gamma ^{2}\left( l+1\right) }%
\right] \left( \cos \frac{\theta }{2}\cos \frac{\theta ^{\prime }}{2}\sin
\frac{\theta }{2}\sin \frac{\theta ^{\prime }}{2}\right) ^{k}
\]
\begin{equation}
\times P_{l-k}^{\left( k,k\right) }(\cos \theta )P_{l-k}^{\left( k,k\right)
}(\cos \theta ^{\prime })e^{ik\left( \varphi -\varphi ^{\prime }\right) }.
\label{a11}
\end{equation}
To include the nonintegrable phase factor due to the AB effect, we
can rename the index $l$ into $q$ related by the relation $l-k=q$.
As a result, Eq. (\ref{a3}) can be written as
\[
\left\{ E-\sum_{q=0}^{\infty }\sum_{k=-\infty }^{\infty }\left[ -\frac{\hbar
^{2}}{2m}\left( \frac{d^{2}}{dr^{2}}+\frac{2}{r}\frac{d}{dr}\right) +\frac{%
(q+k)(q+k+1)\hbar ^{2}}{2mr^{2}}\right] -V(r)\right\}
\]
\[
\times G_{q+k}^{0}(r{\bf ,}r^{\prime };E)\left[ \frac{2(q+k)+1}{4\pi }\frac{%
\Gamma \left( q+1\right) \Gamma \left( q+2k+1\right) }{\Gamma ^{2}\left(
q+k+1\right) }\right] \left( \cos \frac{\theta }{2}\cos \frac{\theta
^{\prime }}{2}\sin \frac{\theta }{2}\sin \frac{\theta ^{\prime }}{2}\right)
^{k}
\]
\begin{equation}
\times P_{q}^{\left( k,k\right) }(\cos \theta )P_{q}^{\left( k,k\right)
}(\cos \theta ^{\prime })e^{ik\left( \varphi -\varphi ^{\prime }\right) }.
\label{a12}
\end{equation}
In addition, the nonintegrable phase factor $\exp \{-i\mu
_{0}\left( 2n\pi \right) \}$ can thus be included with the help of
the Poisson's summation formula (p.124, \cite{18})
\begin{equation}
\sum_{k=-\infty }^{\infty }f(k)=\int_{-\infty }^{\infty }dy\sum_{n=-\infty
}^{\infty }e^{2\pi nyi}f(y).  \label{a13}
\end{equation}
Hence, the expression (\ref{a12}) can be written as
\[
\left\{ E-\sum_{q=0}^{\infty }\int dz\sum_{k=-\infty }^{\infty }\left[ -%
\frac{\hbar ^{2}}{2m}\left( \frac{d^{2}}{dr^{2}}+\frac{2}{r}\frac{d}{dr}%
\right) +\frac{(q+z)(q+z+1)\hbar ^{2}}{2mr^{2}}\right] -V(r)\right\}
\]
\[
\times G_{q+z}(r{\bf ,}r^{\prime };E)\left[ \frac{2(q+z)+1}{4\pi }\frac{%
\Gamma \left( q+1\right) \Gamma \left( q+2z+1\right) }{\Gamma ^{2}\left(
q+z+1\right) }\right] \left( \cos \frac{\theta }{2}\cos \frac{\theta
^{\prime }}{2}\sin \frac{\theta }{2}\sin \frac{\theta ^{\prime }}{2}\right)
^{z}
\]
\begin{equation}
\times P_{q}^{\left( z,z\right) }(\cos \theta )P_{q}^{\left(
z,z\right) }(\cos \theta ^{\prime })e^{i(z-\mu _{0})\left( \varphi
+2k\pi -\varphi ^{\prime }\right) }.  \label{a14}
\end{equation}
Here the superscript $0$ in $G_{q+k}^{0}$ has been suppressed to
reflect
the inclusion of the AB effect. The summation over all indices $k$ forces $%
z=\mu _{0}$ modulo an arbitrary integral number. Therefore, one has
\[
\left\{ E-\sum_{q=0}^{\infty }\sum_{k=-\infty }^{\infty }\left[ -\frac{\hbar
^{2}}{2m}\left( \frac{d^{2}}{dr^{2}}+\frac{2}{r}\frac{d}{dr}\right) +\frac{%
(q+\left| k+\mu _{0}\right| )(q+\left| k+\mu _{0}\right| +1)\hbar ^{2}}{%
2mr^{2}}\right] -V(r)\right\}
\]
\[
\times G_{q+\left| k+\mu _{0}\right| }(r{\bf ,}r^{\prime };E)\left\{ \frac{%
\left[ 2\left( q+\left| k+\mu _{0}\right| \right) +1\right] }{4\pi }\frac{%
\Gamma \left( q+1\right) \Gamma \left( 2\left| k+\mu _{0}\right| +q+1\right)
}{\Gamma ^{2}\left( \left| k+\mu _{0}\right| +q+1\right) }\right\}
e^{ik\left( \varphi -\varphi ^{\prime }\right) }
\]
\begin{equation}
\times \left( \cos \theta /2\cos \theta ^{\prime }/2\sin \theta /2\sin
\theta ^{\prime }/2\right) ^{\left| k+\mu _{0}\right| }P_{q}^{\left( \left|
k+\mu _{0}\right| ,\left| k+\mu _{0}\right| \right) }(\cos \theta
)P_{q}^{\left( \left| k+\mu _{0}\right| ,\left| k+\mu _{0}\right| \right)
}(\cos \theta ^{\prime }).  \label{a15}
\end{equation}
Note that the effect of the AB flux to the radial Green's function
is
to replace the integer quantum number $l$ with the fractional quantum number $%
q+\left| k+\mu _{0}\right| $. Analogously the same procedure can be applied
to the delta function $\delta ^{3}({\bf r-r}^{\prime })$ in the rhs of the Eq. (%
\ref{a1}) with the help of the following solid angle representation of the $%
\delta $ function:
\begin{equation}
\delta \left( \Omega -\Omega ^{\prime }\right) =\sum_{l=0}^{\infty
}\sum_{k=-l}^{l}Y_{lk}(\theta ,\varphi )Y_{lk}^{\ast }(\theta ^{\prime
},\varphi ^{\prime }).  \label{a161}
\end{equation}
Therefore, for the set of the fixed quantum numbers $(q,k)$ one can show
that the radial Green's function satisfies
\[
\left\{ E-\left[ -\frac{\hbar ^{2}}{2m}\left( \frac{d^{2}}{dr^{2}}+\frac{2}{r%
}\frac{d}{dr}\right) +\frac{(q+\left| k+\mu _{0}\right| )(q+\left| k+\mu
_{0}\right| +1)\hbar ^{2}}{2mr^{2}}\right] -V(r)\right\}
\]
\begin{equation}
\times G_{q+\left| k+\mu _{0}\right| }(r{\bf ,}r^{\prime };E)=\delta (r{\bf -%
}r^{\prime }).  \label{a16}
\end{equation}
Hence, the corresponding radial wave equation reads
\begin{equation}
\frac{\hbar ^{2}}{2m}\frac{d^{2}}{dr^{2}}u_{\alpha }(r)+\left[ E-\left( V(r)+%
\frac{\hbar ^{2}}{2m}\frac{\alpha (\alpha +1)}{r^{2}}\right) \right]
u_{\alpha }(r)=0,  \label{a17}
\end{equation}
where we have set $\alpha =q+\left| k+\mu _{0}\right| $, and
$u_{\alpha }(r)=rR_{n\alpha }(r)$. It is clear that $R_{n\alpha }$
satisfies the spherical Bessel equation
\begin{equation}
\left[ \frac{d^{2}}{dr^{2}}+\frac{2}{r}\frac{d}{dr}+\left( \kappa ^{2}-U(r)-%
\frac{\alpha (\alpha +1)}{r^{2}}\right) \right] R_{n\alpha }(r)=0
\label{a18}
\end{equation}
with the definition $\kappa =\sqrt{2mE/\hbar ^{2}}$ and the
reduced
potential $U(r)=2mV(r)/\hbar ^{2}$. For simplicity, we have written $%
R_{n\alpha }(r)$ instead of $R_{n,q,k}(r)$ in which each set
$(n,q,k)$ denote a quantum state. Hence the AB effect reflects
itself by the coupling to the angular momentum in the radial
Green's function, which turns the integer
quantum number to a fractional one.

In order to derive the expectation value $%
\left\langle r^{\lambda }\right\rangle $ in an arbitrary quantum
state $(n,q,k)$ of the ABC
system, one notes that the attractive potential of the Coulomb field is given by $%
V(r)=-Ze^{2}/r$ in which $Ze$ is the total charge at the center of
source and $-e$ representing the charge of the electron. One can
show that the exact solution of the energy spectra in this case is
given by \cite{19,20}
\begin{equation}
E_{n,q,k}=-\frac{Z^{2}e^{2}}{2a_{0}\left[ n+q+\left| k+\mu _{0}\right| +1%
\right] ^{2}},  \label{a19}
\end{equation}
where $a_{0}=\hbar ^{2}/me^{2}$ is the Bohr radius, and the ranges
of the quantum number are $n,q=0,1,2,\cdots ,$ and $-\infty
<k<\infty $. With the help of the result, Eq. (\ref{a17}) can be
brought to the following form:
\begin{equation}
\frac{d^{2}}{dr^{2}}u(r)+\left[ \frac{2Z}{a_{0}r}-\frac{\alpha (\alpha +1)}{%
r^{2}}-\left( \frac{Z}{\tilde{n}a_{0}}\right) ^{2}\right] u(r)=0.
\label{a20}
\end{equation}

\section{Recurrence Formula for the Expectation Value of the Radial operator
of the Aharonov-Bohm-Coulomb System}

For simplicity, we have written $u(r)$ instead of $u_{\alpha }(r)$, and $%
\tilde{n} = \left[ n+q+\left| k+\mu _{0}\right| +1\right] $. For
our purpose of calculating the diagonal matrix element, we will multiply $%
r^{\lambda }u$ to both sides of the above equation. Integrating
this equation with respect to $r$, $ \int_{0}^{\infty }\cdots dr$,
one derives
\begin{equation}
\int_{0}^{\infty }r^{\lambda }u\left( \frac{d^{2}}{dr^{2}}u\right) dr-\alpha
(\alpha +1)\left\langle r^{\lambda -2}\right\rangle +\frac{2Z}{a_{0}}%
\left\langle r^{\lambda -1}\right\rangle -\left( \frac{Z}{\tilde{n}a_{0}}%
\right) ^{2}\left\langle r^{\lambda }\right\rangle =0.  \label{a21}
\end{equation}
With the help of integration by part, the first term yields
\[
\int_{0}^{\infty }r^{\lambda }u\left( \frac{d^{2}}{dr^{2}}u\right) dr=\left.
r^{\lambda }u\left( \frac{d}{dr}u\right) \right| _{0}^{\infty
}-\int_{0}^{\infty }\left( r^{\lambda }\frac{d}{dr}u+\lambda r^{\lambda
-1}u\right) \left( \frac{d}{dr}u\right) dr
\]
\begin{equation}
=\left. \left[ r^{\lambda }u\left( \frac{d}{dr}u\right) -\frac{\lambda }{2}%
r^{\lambda -1}u^{2}\right] \right| _{0}^{\infty }+\frac{\lambda (\lambda -1)%
}{2}\left\langle r^{\lambda -2}\right\rangle -\int_{0}^{\infty }r^{\lambda
}\left( \frac{d}{dr}u\right) ^{2}dr.  \label{a22}
\end{equation}
One can show that the asymptotic behavior of the wave function is
given by \cite{17,19b}
\begin{equation}
\begin{array}{l}
r\longrightarrow 0,\quad u\sim r^{\alpha +1} \\
r\longrightarrow \infty ,\quad u\sim r^{\tilde{n}}e^{-Zr/\tilde{n}a_{0}}
\end{array}
  \label{a23}
\end{equation}
directly from the asymptotic wave equation.
Therefore, one has the necessary and sufficient condition to guarantee the vanishing of the first part of the Eq. (\ref{a22}%
)
\begin{equation}
\left. r^{\lambda }u\left( \frac{d}{dr}u\right) \right| _{0}^{\infty
}=0,\quad \left. r^{\lambda -1}u^{2}\right| _{0}^{\infty }=0,  \label{a24}
\end{equation}
as long as $\lambda >-\left( 2\alpha +1\right) $. Therefore
\begin{equation}
\left[ \frac{\lambda (\lambda
-1)}{2}-\alpha (\alpha +1)\right] \left\langle r^{\lambda
-2}\right\rangle +\frac{2Z}{a_{0}}\left\langle r^{\lambda
-1}\right\rangle -\left( \frac{Z}{\tilde{n}a_{0}}\right)
^{2}\left\langle r^{\lambda }\right\rangle =
\int_{0}^{\infty }r^{\lambda }\left( \frac{d}{dr}%
u\right) ^{2}dr.  \label{a26}
\end{equation}
Similarly, let's multiply $%
2r^{\lambda +1}du/dr$ to the both sides of the Eq. (\ref{a20}) and
then integrate over $\int_{0}^{\infty }\cdots dr$. One can thus
derive the following equations:
\begin{eqnarray}
\int_{0}^{\infty }2r^{\lambda +1}\left( \frac{d}{dr}u\right) \left( \frac{%
d^{2}}{dr^{2}}u\right) dr &=& \left. r^{\lambda +1}\left(
\frac{d}{dr}u\right) ^{2}\right| _{0}^{\infty }-\int_{0}^{\infty
}(\lambda +1)r^{\lambda }\left(
\frac{d}{dr}u\right) ^{2}dr, \label{a27} \\
\int_{0}^{\infty }2r^{\lambda +1}u\left( \frac{d}{dr}u\right) dr
&=& \left. r^{\lambda +1}u^{2}\right| _{0}^{\infty }-(\lambda
+1)\left\langle r^{\lambda }\right\rangle . \label{a271}
\end{eqnarray}
One can also derive similar equations with the power factor
$\lambda+1$ replaced by $\lambda$ or $\lambda-1$ in above
equation. According to the equation (\ref{a24}), the first terms
in the rhs of the Eq.s (\ref{a27})-(\ref{a271}) all vanishes.
Hence, we obtain
\[
(\lambda -1)\alpha (\alpha +1)\left\langle r^{\lambda -2}\right\rangle
-2\lambda \frac{Z}{a_{0}}\left\langle r^{\lambda -1}\right\rangle +(\lambda
+1)\left( \frac{Z}{\tilde{n}a_{0}}\right) ^{2}\left\langle r^{\lambda
}\right\rangle
\]
\begin{equation}
=(\lambda +1)\int_{0}^{\infty }r^{\lambda }\left( \frac{d}{dr}u\right)
^{2}dr.  \label{a28}
\end{equation}
Eliminating the term on the rhs of the above equation with the help of the equation (\ref{a26}),
we finally obtain the general recurrence formula for the ABC system
\begin{equation}
\frac{\lambda +1}{\tilde{n}^{2}}\left\langle r^{\lambda }\right\rangle
-(2\lambda +1)\frac{a_{0}}{Z}\left\langle r^{\lambda -1}\right\rangle +\frac{%
\lambda }{4}\left[ \left( 2\alpha +1\right) ^{2}-\lambda ^{2}\right] \left(
\frac{a_{0}}{Z}\right) ^{2}\left\langle r^{\lambda -2}\right\rangle =0.
\label{a29}
\end{equation}
This formula provides a very convenient tool to calculate the
expectation value of the radial operator $\hat{r}^{\lambda }$ for
arbitrary power $\lambda $ in any arbitrary quantum state
$(n,q,k)$ for the ABC system without facing the complication
computing the wave functions directly. For example, let's
calculate a few leading terms of the expectation value of the
radial operator. First of all, one can set $\lambda =0$ where the
recurrence formula (\ref{a29}) gives
\begin{equation}
\left\langle \frac{1}{r}\right\rangle _{n,q,k}=\frac{Z}{\tilde{n}^{2}a_{0}}=%
\frac{Z}{\left( n+q+\left| k+\mu _{0}\right| +1\right) ^{2}a_{0}}.
\label{a30}
\end{equation}
Note that the effect of the magnetic flux leads to the decrease of
the expectation value of the potential. Note that this effect has
no correspondence in classical system. Moreover, if we choose
$\lambda =1,2$, one will instead derive
\begin{equation}
\begin{array}{l}
\left\langle r\right\rangle _{n,q,k}=\frac{1}{2}\left[ 3\tilde{n}^{2}-\alpha
(\alpha +1)\right] \frac{a_{0}}{Z}, \\
\left\langle r^{2}\right\rangle _{n,q,k}=\frac{\tilde{n}^{2}}{2}\left[ 1+5%
\tilde{n}^{2}-3\alpha (\alpha +1)\right] \left( \frac{a_{0}}{Z}\right) ^{2}.
\end{array}
\label{a31}
\end{equation}
These results show that the magnetic flux effect will increase the expectation of the radial
operator such as the case where the quantum states given by $(n,q,k)=(0,0,0)$. Indeed, one can show that
\begin{equation}
\begin{array}{l}
\left\langle r\right\rangle _{0,0,0}=\frac{1}{2}\left[ 2\left| \mu
_{0}\right| ^{2}+5\left| \mu _{0}\right| +3\right] \frac{a_{0}}{Z}, \\
\left\langle r^{2}\right\rangle _{0,0,0}=\frac{(\left| \mu _{0}\right|
+1)^{2}}{2}\left[ 2\left| \mu _{0}\right| ^{2}+7\left| \mu _{0}\right| +6%
\right] \left( \frac{a_{0}}{Z}\right) ^{2}.
\end{array}
\label{a32}
\end{equation}
Note that these results reduce to the well-known pure Coulomb system, where $\mu_0=0$, $%
\left\langle r\right\rangle _{0,0,0}=3a_{0}/2Z$, and $\left\langle
r^{2}\right\rangle _{0,0,0}=3(a_{0}/Z)^{2}$ where the nonlocal
effect of the magnetic flux in the physical average becomes
manifest in these formulae.

According to the definition leading to the Eq. (\ref{a20}), we
have $\alpha _{\max }=\left( q+\left| k+\mu _{0}\right| \right)
_{\max }=\tilde{n}-1$ corresponding to the nodes of the wave
function at $n=0$ for the modified circular Bohr orbit. In this
case, the radial wave function reduces to \cite{17}
\begin{equation}
u_{0,q,k}(r)=rR_{0,q,k}(r)=r^{\tilde{n}}e^{-Zr/\tilde{n}a_{0}}.  \label{c1}
\end{equation}
Hence we can calculate the most possible position of the radius $r_{{\rm most}}$, defined as the position where the
probability function $|u|^2$ attains its maximal, given by $%
du_{0,q,k}(r=r_{{\rm most}})/dr=0$. The result indicates that
\begin{equation}
r_{{\rm most}}=\frac{\tilde{n}^{2}a_{0}}{Z}  \label{c2}
\end{equation}
for the most possible position of the radius. It is clear that the
presence of the magnetic flux increases the value of the most
possible position. On the other hand, Eq. (\ref{a31}) gives the
following radial expectation value,
when $\alpha $ is equal to $\alpha _{\max }$,%
\begin{equation}
\left\langle r\right\rangle _{0,\tilde{n}-1}=\left[ \tilde{n}^{2}+\frac{%
\tilde{n}}{2}\right] \frac{a_{0}}{Z}.  \label{c3}
\end{equation}
Therefore quantum mechanical average (the expectation value) for
the radius of the modified Bohr orbit is bigger then the most
possible position of the radius. In addition, the position
fluctuation is given by
\begin{equation}
\left( \triangle r\right) _{0,\tilde{n}-1}=\sqrt{\left\langle
r^{2}\right\rangle _{0,\tilde{n}-1}-\left\langle r\right\rangle _{0,\tilde{n}%
-1}^{2}}=\sqrt{\left[ \frac{\tilde{n}^{3}}{2}+\frac{\tilde{n}}{4}^{2}\right]
}\frac{a_{0}}{Z}  \label{c4}
\end{equation}
which implies the increase of the quantum fluctuation due to the
increase of the magnetic flux. Consequently, one can derive the
ratio between the fluctuation and the corresponding average
\begin{equation}
\frac{\left( \triangle r\right) _{0,\tilde{n}-1}}{\left\langle
r\right\rangle _{0,\tilde{n}-1}}=\frac{1}{\sqrt{2\tilde{n}+1}}  \label{c5}
\end{equation}
which indicates that the quantum mechanical result tends to
approach the Bohr picture of orbit quantization as $\tilde{n}$
increases or equivalently the magnetic flux increases.

If we choose $\lambda =-1,-2$, the recurrence relation indicates
that
\begin{equation}
\left\langle r^{-3}\right\rangle _{n,q,k}=\frac{a_{0}/Z}{\alpha (\alpha
+1)(a_{0}/Z)^{2}}\left\langle r^{-2}\right\rangle _{n,q,k}=\frac{1}{\tilde{n}%
^{3}\alpha (\alpha +1/2)(\alpha +1)}\left( \frac{Z}{a_{0}}\right) ^{3},
\label{a33}
\end{equation}
\begin{equation}
\left\langle r^{-4}\right\rangle _{n,q,k}=\frac{3\tilde{n}^{2}-\alpha
(\alpha +1)}{2\tilde{n}^{5}(\alpha -1/2)(\alpha +1/2)(\alpha +1)(\alpha +3/2)%
}\left( \frac{Z}{a_{0}}\right) ^{3}.  \label{a34}
\end{equation}
Here we have used the result
\begin{equation}
\left\langle r^{-2}\right\rangle _{n,q,k}=\frac{1}{\tilde{n}^{3}(\alpha +1/2)%
}\left( \frac{Z}{a_{0}}\right) ^{2},  \label{a35}
\end{equation}
which is the only moment can not be derived from the recurrence
Eq. (\ref{a29}). It can, however, be derived from the well-known
Hellmann-Feynman formula
\begin{equation}
\frac{\partial E_{n,q,k}}{\partial \alpha }=\left\langle \Psi _{n,q,k}\left|
\frac{\partial \hat{H}}{\partial \alpha }\right| \Psi _{n,q,k}\right\rangle ,
\label{b5}
\end{equation}
where the Hamiltonian operator is given by
\begin{equation}
\hat{H}=-\frac{\hbar ^{2}}{2m}\frac{d^{2}}{dr^{2}}+\frac{\hbar ^{2}}{2m}%
\frac{\alpha (\alpha +1)}{r^{2}}-\frac{Ze^{2}}{r}.  \label{b6}
\end{equation}
It hence follows that
\begin{equation}
\frac{\partial E_{n,q,k}}{\partial \alpha }=(\alpha +\frac{1}{2})\frac{\hbar
^{2}}{m}\left\langle r^{-2}\right\rangle _{n,q,k}.  \label{b7}
\end{equation}
With the help of the Eq. (\ref{a19}), one derives the result shown
in Eq. (\ref{a35}). This result gives us the ratio information of
the modified centrifugal potential in quantum mechanics, namely,
\begin{equation}
\left\langle \frac{\alpha (\alpha +1)\hbar ^{2}}{2mr^{2}}\right\rangle
_{n,q,k}=-\frac{\alpha (\alpha +1)}{(\alpha +1/2)\tilde{n}}E_{\tilde{n}}.
\label{b8}
\end{equation}
Since the total kinetic energy of the system equals $
<\tilde{T}>_{n,q,k}=-E_{\tilde{n}}$ (see appendix for
details), the ratio of the centrifugal potential to the total
kinetic energy, $r_c=<V_c>/<\tilde{T}>$, is
$$r_c=\alpha (\alpha
+1)/\left[ (\alpha +1/2)\tilde{n}\right] . $$
This result
indicates that, for fixed $\tilde{n}$, the increase of the
intensity of the magnetic flux induces the increase of the ratio
$r_c$ and thus the weight of the centrifugal potential.
When $\alpha $ is equal to $\alpha _{\max }=\tilde{%
n}-1$ corresponding to the modified Bohr orbit case, the ratio
$r_c$ become
\begin{equation}
r_c \longrightarrow \frac{%
\tilde{n}-1}{\tilde{n}-1/2}.  \label{b9}
\end{equation}
On the other hand, the ratio of the average of the radial kinetic
energy to the total kinetic energy, $r_r = <p_r^2/2m> / <
\tilde{T} >$, is given by
\begin{equation}
r_r = \left\langle \frac{p_{r}^{2}}{2m}\right\rangle _{0,q,k}=\frac{1}{2\tilde{n}-1%
}  \label{b10}
\end{equation}
at the limit $\alpha = \alpha_{max}$. Therefore when $\tilde{n}\gg
1$, the radial kinetic energy will become very small reproducing
the classical result.

In addition, one can also discuss the effect of the magnetic flux
in the well-known formulae introduced  by J. Schwinger \cite{19a}.
It was shown that
\begin{equation}
\begin{array}{l}
\left\langle \frac{dV}{dr}\right\rangle =\frac{2\pi \hbar ^{2}}{m}\left|
\Psi (0)\right| ^{2},\quad l=0 \\
\left\langle \frac{dV}{dr}\right\rangle =\left\langle \frac{{\bf l}^{2}}{%
mr^{3}}\right\rangle =l(l+1)\frac{\hbar ^{2}}{m}\left\langle \frac{1}{r^{3}}%
\right\rangle ,\quad l\neq 0.
\end{array}
\label{b11}
\end{equation}
These results are very useful and of considerably general interest
for the study of the bound states in a system with a central
potential. The first identity relates the $s$-wave wave function
at the origin to the gradient of the potential. This identity
comes explicitly with the Planck constant $\hbar $ indicating the
significance of the quantum effect. Therefore there is no
classical correspondence. The second identity is, however, a
generalization of the quantum effect of an important classical
theory
\begin{equation}
\left( \frac{dV}{dr}\right) _{{\rm av}}=\frac{{\bf l}^{2}}{m}\left( \frac{1%
}{r^{3}}\right) _{{\rm av}},  \label{b12}
\end{equation}
where $\left( f \right) _{{\rm av}}$ represents the average of the
periodic function $f(t)$ over a classical period. Eq. (\ref{b12})
states that the average of a external force equals to the average
of the centripetal force over the classical period.

In the presence of the magnetic flux, the Hamilton operator is
given by (\ref {a18})
\begin{equation}
\hat{H}=-\frac{\hbar ^{2}}{2m}\left( \frac{\partial ^{2}}{\partial r^{2}}+%
\frac{2}{r}\frac{\partial }{\partial r}\right) +\frac{\alpha (\alpha
+1)\hbar ^{2}}{2mr^{2}}+V(r).  \label{b13}
\end{equation}
Computing the commutator of the operators $\partial /\partial r $
and $\hat{H}$, one has
\begin{equation}
\left[ \frac{\partial }{\partial r},\hat{H}\right] =\frac{\hbar ^{2}}{mr^{2}}%
\frac{\partial }{\partial r}-\frac{\alpha (\alpha +1)\hbar ^{2}}{mr^{3}}+%
\frac{dV}{dr}.  \label{b14}
\end{equation}
Evaluating the expectation value with respect to an arbitrary
bound state which makes the contribution from the lhs vanished, we
would have
\[
\left\langle \frac{dV}{dr}\right\rangle -\frac{\alpha (\alpha +1)\hbar ^{2}}{%
m}\left\langle \frac{1}{r^{3}}\right\rangle =-\frac{\hbar ^{2}}{m}%
\left\langle \frac{1}{r^{2}}\frac{\partial }{\partial r}\right\rangle
\]
\begin{equation}
=-\frac{\hbar ^{2}}{m}\int r^{2}drd\Omega \Psi ^{\ast }\frac{1}{r^{2}}\frac{%
\partial }{\partial r}\Psi ,  \label{b15}
\end{equation}
with $d\Omega $ the solid angle. Since the mean value in the lhs
of the above equation is a real number, so does the rhs of the
above equation. Hence one has
\begin{equation}
\int dr\Psi ^{\ast }\frac{\partial }{\partial r}\Psi =\int dr\Psi \frac{%
\partial }{\partial r}\Psi ^{\ast }=\frac{1}{2}\int dr\frac{\partial }{%
\partial r}\left( \Psi \Psi ^{\ast }\right) =\left. \frac{1}{2}\Psi \Psi
^{\ast }\right| _{0}^{\infty }=0.  \label{b16}
\end{equation}
In deriving the above equations, we have used the following facts:
(1) the wave function of the bound state vanishes at the region
$r\rightarrow \infty $; (2) in the presence of the magnetic flux,
the vanishing of the wave function $\Psi (0)=0$ holds at the
origin \cite{19b}. To prove the second proposition, one notes that
the asymptotic form of the wave function can be shown to be
\cite{19b}
\begin{equation}
u_{\alpha }(r)\sim r^{\alpha +1},\quad \alpha =q+\left| k+\mu _{0}\right| >0,
\label{b17}
\end{equation}
near the origin in the presence of the magnetic flux. Because that
$u_{\alpha }(r)=rR_{\alpha }(r)$, one has immediately that
$u_{\alpha }^{\prime
}(r)=R_{\alpha }(r)+rR_{\alpha }^{\prime }(r)$.
Therefore we obtain $%
u_{\alpha }^{\prime }(0)=R_{\alpha }(0)$. In addition, Eq. (\ref
{b17}) says that $u_{\alpha }^{\prime }(r)=(\alpha +1)r^{\alpha }$ implying that $%
u_{\alpha }^{\prime }(0)=0$. This proves the second proposition
that $\Psi (0)=0$.

Accordingly, in a central force system with a magnetic flux, one
can generalize the Schwinger theorem  (\ref{b11}) to the following
form:
\begin{equation}
\left\langle \frac{dV}{dr}\right\rangle _{n,q,k}=\frac{\alpha (\alpha
+1)\hbar ^{2}}{m}\left\langle \frac{1}{r^{3}}\right\rangle _{n,q,k}.
\label{b18}
\end{equation}
In addition, the $s$-wave no longer exists in the presence of the
magnetic flux, there is no corresponding generalization of the
first identity in the Eq. (\ref{b11}). On the other hand, we find
that the topological effect of the magnetic flux has a generalized
quantum mechanical shown in Eq. (\ref{b11}) although such effect
does not have any dynamical significance in classical mechanics.

\section{Conclusion}

We have derived a general recurrence formula of the expectation
value of the radial operator with different power for an
Aharonov-Bohm-Coulomb system providing us a convenient method to
compute the average of the radial operator in any power. With this
useful recurrence formula, one can avoid tedious and complicate
integration with the wave functions. The useful formulae and
techniques shown in this paper have many interesting applications
in the study of many physical systems in the presence of a
magnetic field. For example, we have shown the generalized theorem
of Schwinger for bound state system under the influence of a
magnetic field.

In addition, the two dimensional (2D) system is important in the
fractional quantum Hall effect and the high-$T_{c}$
superconductivity, we will also generalize our result to the 2D
system at the end of this paper. Two dimensional central force
quantum system is governed by the radial Schr\"{o}dinger equation
\cite{22}
\begin{equation}
\left\{ -\frac{\hbar ^{2}}{2m}\left[ \left( \frac{d^{2}}{d\rho ^{2}}+\frac{1%
}{\rho }\frac{d}{d\rho }\right) -\frac{k^{2}}{\rho ^{2}}\right] +V(\rho
)\right\} R_{nk}=ER_{nk},  \label{e1}
\end{equation}
where $\rho $ is the corresponding radial length in 2D, i.e. $\rho
=\sqrt{x^{2}+y^{2}}$. The corresponding total wave functions is
given by $\Psi (\rho ,\varphi )=R_{nk}(\rho )e^{ik\varphi }$ with
the range of the quantum number specified by $k=0,\pm 1,\pm
2,\cdots $. In addition, the energy spectra $E$ depends on $\left|
k\right| $ and the wave function $R_{n\left| k\right| }$
corresponding to the energy level $E_{n\left| k\right| }$
satisfies the orthonormal condition
\begin{equation}
\int_{0}^{\infty }d\rho \rho R_{n\left| k\right| }^{\ast }(\rho
)R_{n^{\prime }\left| k\right| }(\rho )=\delta _{nn^{\prime }}.  \label{e2}
\end{equation}
If we perform a transformation $R_{n\left| k\right| }(\rho )=\chi _{n\left|
k\right| }(\rho )/\sqrt{\rho }$, Eq. (\ref{e1}) will reduce to
\begin{equation}
\left\{ -\frac{\hbar ^{2}}{2m}\left[ \frac{d^{2}}{d\rho ^{2}}+\frac{%
(k-1/2)(k+1/2)}{\rho ^{2}}\right] +V(\rho )\right\} \chi _{n\left| k\right|
}(\rho )=E\chi _{n\left| k\right| }(\rho ).  \label{e3}
\end{equation}
The AB effect can be introduced by the similar procedure leading to the Eqs. (\ref{a13})$%
\sim $(\ref{a18}). The result is
\begin{equation}
\left\{ -\frac{\hbar ^{2}}{2m}\left[ \frac{d^{2}}{d\rho ^{2}}+\frac{(\tilde{%
\alpha}-1/2)(\tilde{\alpha}+1/2)}{\rho ^{2}}\right] +V(\rho )\right\} \chi
_{n\left| k\right| }(\rho )=E\chi _{n\left| k\right| }(\rho ),  \label{e4}
\end{equation}
where we have defined $\tilde{\alpha}=\left| k+\mu _{0}\right| $.
Comparing this equation with Eq. (\ref{a17}), one finds that
results derived for the 3D system can be generalized to the 2D
system with the parameters $\alpha$ and $\tilde {\alpha}$ related
by the following relation
\begin{equation}
\alpha (\alpha +1)\longleftrightarrow (\tilde{\alpha}-1/2)(\tilde{\alpha}%
+1/2),  \label{e5}
\end{equation}
or simply
$$\alpha \longleftrightarrow (\tilde{\alpha}-1/2).$$

For example, we find that the energy spectra of the 2D ABC system
is given by
\begin{equation}
E_{n,\tilde{\alpha}}=-\frac{Z^{2}e^{2}}{2a_{0}\left[ n+\left| k+\mu
_{0}\right| +1/2\right] ^{2}}\equiv -\frac{Z^{2}e^{2}}{2a_{0}\tilde{n}%
_{2}^{2}}.  \label{e6}
\end{equation}
This exact formula agrees with the previous result obtained in
Ref. \cite{20}. In addition, one can also derive some useful and
important expectation values for the 2D ABC system:
\begin{equation}
\left\langle \frac{1}{\rho }\right\rangle _{n,k}=\frac{Z}{\tilde{n}_{2}^{2}}%
a_{0},  \label{e7}
\end{equation}
\begin{equation}
\left\langle \frac{1}{\rho ^{2}}\right\rangle _{n,k}=\frac{1}{\tilde{n}%
_{2}^{3}\tilde{\alpha}}\left( \frac{Z}{a_{0}}\right) ^{2},  \label{e8}
\end{equation}
\begin{equation}
\left\langle \frac{1}{\rho ^{3}}\right\rangle _{n,k}=\frac{1}{\tilde{n}%
_{2}^{3}\tilde{\alpha}(\tilde{\alpha}-1/2)(\tilde{\alpha}+1/2)}\left( \frac{Z%
}{a_{0}}\right) ^{3}.  \label{e9}
\end{equation}
Generalization to many other cases are straightforward following
similar procedures.

\section*{Acknowledgments}

This work is supported in part by the National Science Council
under the grant number NSC90-2112-M009-021.

\section{Appendix}

In the appendix, we shall prove the Virial theorem for the
spherically symmetric system under an AB magnetic flux. The
Hamiltonian for the spherically symmetric system under an AB
magnetic flux is given by the Eq. (\ref{a18})
\begin{equation}
\hat{H}=\frac{\hat{p}_{r}^{2}}{2m}+\frac{\alpha (\alpha +1)\hbar ^{2}}{%
2mr^{2}}+V(r),  \label{d1}
\end{equation}
where the Hermite operator is defined as
\begin{equation}
\hat{p}_{r}=\frac{\hbar }{i}\left( \frac{\partial }{\partial r}+\frac{1}{r}%
\right) =\hat{p}_{r}^{\dagger }.  \label{d2}
\end{equation}
To obtain the Virial theorem, one notes that the Heisenberg
equations of motion for the position and the momentum operators
are
\begin{equation}
\frac{d{\bf r}}{dt}=\frac{1}{i\hbar }\left[ {\bf r,}\hat{H}\right] =\nabla _{%
{\bf p}}\hat{H}=\hat{e}_{p_{r}}\frac{\hat{p}_{r}}{m},  \label{d3}
\end{equation}

\begin{equation}
\frac{d{\bf p}}{dt}=\frac{1}{i\hbar }\left[ {\bf p,}\hat{H}\right] =\nabla _{%
{\bf r}}\hat{H}=\left[ \frac{\alpha (\alpha +1)\hbar ^{2}}{mr^{3}}-\frac{%
dV(r)}{dr}\right] \hat{e}_{r},  \label{d4}
\end{equation}
where $\hat{e}_{p_{r}}$ is the unit vector of $p_{r}$ in the
momentum space. Therefore, we have
\[
\frac{d}{dt}\left( {\bf r\cdot p}\right) =\frac{1}{i\hbar }\left[ {\bf %
r\cdot p,}\hat{H}\right]
\]
\begin{equation}
=\frac{d{\bf r}}{dt}\cdot {\bf p+r\cdot }\frac{d{\bf p}}{dt}=\frac{\hat{p}%
_{r}^{2}}{m}+\frac{\alpha (\alpha +1)\hbar ^{2}}{mr^{2}}-r\frac{dV(r)}{dr}.
\label{d5}
\end{equation}
Average with respect to the wave functions, we obtain
\begin{equation}
\left\langle \frac{\hat{p}_{r}^{2}}{2m}+\frac{\alpha (\alpha +1)\hbar ^{2}}{%
2mr^{2}}\right\rangle =\frac{1}{2}\left\langle r\frac{dV(r)}{dr}%
\right\rangle ,  \label{d6}
\end{equation}
where we have used the identity $\left\langle \left[ {\bf r\cdot p,}\hat{H}%
\right] \right\rangle =0$. Let us define the total kinetic energy
operator as
\begin{equation}
\tilde{T}=\frac{\hat{p}_{r}^{2}}{2m}+\frac{\alpha (\alpha +1)\hbar ^{2}}{%
2mr^{2}}.  \label{d7}
\end{equation}
Hence one can derive the following Virial theorem for the
spherically symmetric system under an AB magnetic flux
\begin{equation}
2\left\langle \tilde{T}\right\rangle =\left\langle r\frac{dV(r)}{dr}%
\right\rangle .  \label{d8}
\end{equation}
In particular, if the potential $V(r)$ is a homogeneous function of degree $\nu $%
, namely $V(br)=b^{\nu }V(r)$, the Eq. (\ref{d8}) can be shown to
be
\begin{equation}
2\left\langle \tilde{T}\right\rangle =\nu \left\langle V(r)\right\rangle .
\label{d9}
\end{equation}
Together with the total energy relation,
\begin{equation}
\left\langle \tilde{T}+V\right\rangle =E_{n,q,k},  \label{d10}
\end{equation}
it is easily to show that the relations between the averages of
the kinetic and the potential energy, and hence the total energy
\begin{equation}
\begin{array}{l}
\left\langle \tilde{T}\right\rangle _{n,q,k}=\frac{\nu }{\nu +2}E_{n,q,k},
\\
\left\langle V\right\rangle _{n,q,k}=\frac{2}{\nu +2}E_{n,q,k}.
\end{array}
\label{d11}
\end{equation}
For an ABC system ($\nu =-1$), one has
$<\tilde{T}>_{n,q,k}=-E_{n,q,k}$, and $<V>_{n,q,k}=2 E_{n,q,k}$.

\end{document}